\documentclass[aps,prl,twocolumn,superscriptaddress]{revtex4-1}

\usepackage{graphicx}
\usepackage[rightcaption]{sidecap}
\usepackage{floatrow}
\usepackage{amsmath,mathrsfs}
\usepackage{array}
\usepackage{color}
\newcommand{\ket}[1]{{\left| {#1} \right\rangle}}
\newcommand{\bra}[1]{{\left\langle {#1} \right|}}
\newcommand{\ketbra}[2]{{\left| {#1} \right\rangle
    \!\!\left\langle{#2} \right|}}

\DeclareMathAlphabet{\mathbsf}{OT1}{cmss}{bx}{n}

\newcommand{\boldgreek}[1]{{\mbox{\boldmath$ {#1} $}}}

\begin{document}

%\preprint{}

\title{Dipolar quantum logic for freely-rotating trapped molecular ions}

\author{Eric R. Hudson}
\author{Wesley C. Campbell}
\affiliation{Department of Physics and Astronomy, University of California -- Los Angeles, Los Angeles, California, 90095, USA}

\date{\today}

\begin{abstract}
We consider the practical feasibility of using the direct, electric dipole-dipole interaction between co-trapped molecular ions for robust quantum logic without the need for static polarizing fields. The use of oscillating dipole moments, as opposed to static electric dipoles, dynamically decouples the dipoles from laboratory fields, including the electric fields of the trap itself. Further, this implementation does not require quantum control of motion, potentially removing a major roadblock to ion trap quantum computing scalability. Since the polarizing field is electromagnetic radiation, even pairs of states with splittings in the THz regime can be fully polarized.  
\end{abstract}


\maketitle

%\section{Introduction}
Of all quantum architectures explored to date, trapped atomic ions have demonstrated the lowest single- and two-qubit gate error rates~\cite{Harty2014,Balance2016,Gaebler2016}, as well as single-qubit storage times exceeding 10 minutes~\cite{Wang2017}. 
In fact, fully programmable few-qubit quantum computers based on trapped ions have already been constructed~\cite{Debnath2016}.
These systems rely on the quantum nature of the motion of the trapped ions to produce entanglement between qubits. 
However, an unanticipated phenomenon, dubbed anomalous heating, which causes the trapped ion qubits to spontaneously heat when held near a surface has slowed the scaling of ion trap quantum computers to a large number of qubits~\cite{Sedlacek2018}.

One possible solution to this problem is the use of an entangling interaction that does not rely on the quantum nature of motion, such as the long-range, electric dipole-dipole interaction between polar particles. 
However, as eigenstates of parity cannot support an electric dipole moment, these techniques typically require a polarizing electric field to orient the dipoles in the laboratory frame, leading to a net dipole-dipole interaction between particles.  
In proposals of Jaksch \textit{et al.} for Rydberg atoms \cite{Jaksch2000} and DeMille for polar molecules~\cite{DeMille2002} a static electric field provides this polarization.  
In proposals of Brennen \textit{et al.}~\cite{Brennen1999} and Lukin and Hemmer~\cite{Lukin2000} for co-located neutral atoms and Lukin \textit{et al}.~\cite{Lukin2001} for Rydberg atoms, an optical field is present to enable a net dipolar interaction for particles separated by less than a wavelength.
These techniques are difficult to implement in a traditional ion trap as the trapping dictates that the average electric field experienced by the ion $\langle\vec{E}\rangle = 0$ and the ion-ion Coulomb repulsion prevents high-density samples. 

Here, we describe a technique to effect a dipole-dipole interaction and enable entanglement between nearby trapped polar molecular ions without the use of a continuously applied polarizing electric field.
By creating a superposition of opposite parity eigenstates, the time-dependent electric polarization can mediate a net dipole-dipole interaction.
The underlying spin exchange type interaction has been extensively explored both theoretically and experimentally in the context of quantum simulation with neutral atoms~\cite{Syzranov2016}, neutral molecules~\cite{Micheli2006,Barnett2006,Gorshkov2011,Yao2012,Yan2013,Hazzard2014}, and Rydberg atoms~\cite{Barredo2015,Maghrebi2015}.  
We show that by polarizing the molecules along the axial direction of a linear Paul trap, the strength of the dipole-dipole interaction is maximized while the trap-field-induced decoherence is suppressed. 
We describe a realistic implementation with trapped $\mbox{CO}^+$, which indicates that without cooling to the motional ground state an entanglement fidelity $\geq$~0.9999 is achievable with gate times of $\sim$10-100$~\mu$s and that electric field induced decoherence times can be $\geq10^4$~s.

We consider a system composed of two polar molecular ions, each with a pair of identical, opposite-parity states, $\ket{\mathrm{g}}_i$ and $\ket{\mathrm{e}}_i$ that represent the $-1$ and $+1$ eigenstates, respectively, of the Pauli operator $\sigma_z^{(i)}$ for this effective 2-level system of molecule $i$.  
These states are separated in energy by the non-interacting Hamiltonian $\mathscr{H}_0=\frac{\Delta}{2}(\sigma_z^{(1)} + \sigma_z^{(2)})$ (with $\hbar \equiv 1$) and are connected by an electric dipole transition moment $d = |\bra{\mathrm{g}}\vec{\mathbf{d}}\ket{\mathrm{e}}|$.
We will refer to these states as rotational states of the molecule, but they could be other dipole connected states, e.g. $\Omega$-doublet states. 
Manipulations of the molecules (single-qubit gates) are effected by applied resonant radiation 
%(we will refer to this radiation as microwaves, though below we discuss applications in other frequency regimes).
whose electric field at the position of molecule $i$ is given by $\mathbf{E}^{(i)}(t) = \frac{E^{(i)}}{2}(\hat{\boldgreek{\epsilon}}\,e^{-\imath \omega t} + \hat{\boldgreek{\epsilon}}^\ast e^{\imath \omega t})$.  The molecules interact with the radiation via
\begin{equation}
\mathscr{H}_\mathrm{rad} = \left( \frac{\Omega^{(1)}}{2} \,\sigma_+^{(1)}e^{-\imath \omega t} + \frac{\Omega^{(2)}}{2}\, \sigma_+^{(2)}e^{-\imath \omega t} + \mathrm{H.c.} \right)
\end{equation}
%For two molecules spatially separated by $\vec{\mathbf{r}}$, the evolution of the system is governed (in the rotating wave approximation) by the Hamiltonian
%\begin{eqnarray}
%\mathscr{H} &=& \mathscr{H}_0 + \mathscr{H}_\mathrm{rad} + \mathscr{V}_\mathrm{dd}(\vec{\mathbf{r}}) \\ \nonumber
%\mathscr{H}_\mathrm{rad} &=& \left( \frac{\Omega^{(1)}}{2}
%\,\sigma_+^{(1)}e^{-\imath \omega t} + \frac{\Omega^{(2)}}{2}\,
%\sigma_+^{(2)}e^{-\imath \omega t} + \mathrm{H.c.} \right)
%\end{eqnarray}
where $\sigma_+^{(i)} \equiv {_{}}_i\ketbra{\mathrm{e}}{\mathrm{g}}_i$ and $\Omega^{(i)} \equiv {_{}}_i\bra{\mathrm{g}}\vec{\mathbf{d}}\cdot \mathbf{E}^{(i)} \ket{\mathrm{e}}_i$ is the resonant Rabi frequency at the position of molecule $i$.

The molecules interact via the dipole-dipole interaction, which in the interaction picture (rotating frame) with respect to $\mathscr{H}_0$ takes the form~\cite{Micheli2006}

\begin{equation}
\mathscr{V}_\mathrm{dd}(\vec{\mathbf{r}}) =  \frac{J}{2}(\sigma_x^{(1)}\sigma_x^{(2)} + \sigma_y^{(1)}\sigma_y^{(2)}) \label{XXHamiltonian} 
\end{equation}
where $J(\vec{\mathbf{r}}) = d^2/(4\pi\epsilon_o r^3)(1-3\cos^2\theta)$ is the dipole-dipole interaction strength and $\theta$ is the angle between the polarization axis and $\vec{\mathbf{r}}$. 
We consider only parallel, linearly polarized dipoles for simplicity, and Eq.~\ref{XXHamiltonian} is obtained from the exact interaction by making the rotating wave approximation.
The separation between particles, $\vec{\mathbf{r}}$, is assumed fixed, though this approximation will be relaxed when decoherence from thermal motion is considered. The time evolution due to the dipole-dipole interaction is then:

\begin{align}
U_\mathrm{dd}(t)  = &\mathbf{1}+ \left(\cos{\left(J t\right)}-1\right)\left(\ket{\mathrm{g},\mathrm{e}}\bra{\mathrm{g},\mathrm{e}}
 + \ket{\mathrm{e},\mathrm{g}}\bra{\mathrm{e},\mathrm{g}}\right)\\ \nonumber 
& -\imath\sin{\left(J t\right)}\left(\ket{\mathrm{g},\mathrm{e}}\bra{\mathrm{e},\mathrm{g}} + \ket{\mathrm{e},\mathrm{g}}\bra{\mathrm{g},\mathrm{e}}\right)
\label{XXEvolution} 
\end{align}

\begin{figure}[t]
\includegraphics[width=1\textwidth]{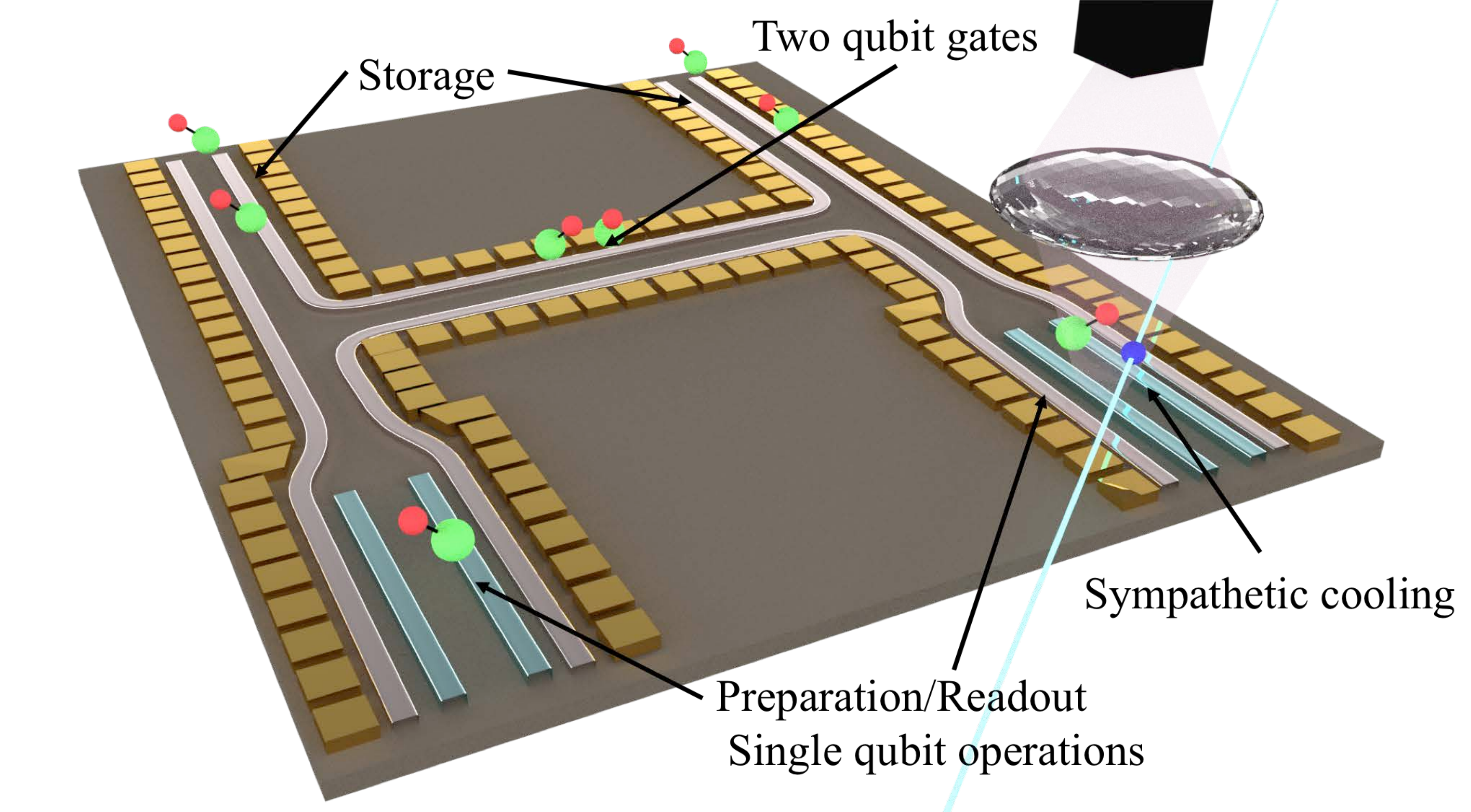}
\caption{Schematic of chip-based molecular ion quantum processor. Separate regions can be optimized for qubit state-preparation  and read-out, two-qubit gates, and qubit storage. Because the implementation does not use the motional modes for quantum information processing, the anomalous heating observed in chip traps~\cite{Sedlacek2018} is not a limitation to its scalability. This allows continuous sympathetic cooling of the molecular ions via Doppler cooled atomic ions.}
\label{fig:schematic}
\end{figure}

A possible physical implementation, based on surface electrode trapping~\cite{Brown2016}, for realizing this system with molecular ions is sketched in Fig.~\ref{fig:schematic}.  
State preparation and/or readout of the molecular ion qubits could be accomplished with techniques including, sympathetic cooling~\cite{Rellergert2013,Hansen2014,Hauser2015}, coupling to microwave resonators~\cite{Schuster2011}, state-dependent heating~\cite{Patterson2018}, quantum logic spectroscopy~\cite{Wolf2016, Chou2017}, and optical cycling~\cite{Lien2014}. 
Two-qubit gates, described next, could occur in a separate region accessed by shuttling~\cite{Blakestad2009}. 
Doppler cooling of atomic ions can be used to continuously sympathetically cool the molecular ion qubits. 

The basic scheme for performing quantum logic operations can be illustrated by considering the specific example of creating a Bell state of two trapped polar molecular ions.
Here we briefly sketch two ways in which this can be accomplished using either local or global single-qubit gates. 

First, for experimental implementations where high fidelity molecule-resolved single-qubit gates are available, for two molecules initially in $\ket{\psi_1,\psi_2}\! =\!  \ket{\mathrm{g},\mathrm{g}}$, a fast ($\tau \ll 1/J$) $\pi$-pulse addressed to molecule 1 will prepare $\ket{\psi}=\ket{\mathrm{e},\mathrm{g}}$.
After the pulse, the molecules evolve freely in time under the influence of Eq.~\ref{XXHamiltonian}.
After a time $t_\mathrm{G} = \frac{\pi}{4J}$, the system has evolved to the entangled state $\ket{\psi^\prime} = \frac{1}{\sqrt{2}} \left( \ket{\mathrm{e},\mathrm{g}} - \imath \ket{\mathrm{g},\mathrm{e}} \right)$.  Another fast $\pi$-pulse on molecule 1 and a $\sigma_z^{(1)}$ rotation of molecule 1 can yield the Bell state $\ket{\psi_\mathrm{f}} = \frac{1}{\sqrt{2}}\left( \ket{\mathrm{e},\mathrm{e}}+ e^{\imath \phi}\ket{\mathrm{g},\mathrm{g}}\right)$ where $\phi$ is chosen by the phase of the final two rotations.  
Since the state $\ket{\psi_\mathrm{f}}$ is an eigenstate of Eq.~\ref{XXHamiltonian}, this stops the dynamics and the Bell state can be stored or used as a resource in further experiments. 
In the architecture of Fig.~\ref{fig:schematic}, this procedure could be accomplished by, for example, preparing the ions separately in the state preparation zones and shuttling them together for interaction. 

Second, if we restrict ourselves to global operations, for two molecules initially in $\ket{\psi_1,\psi_2}\! =\!  \ket{\mathrm{g},\mathrm{g}}$, a fast, global, resonant $\frac{\pi}{2}$-pulse
%polarized along $\vec{r}$
can prepare $\ket{\psi}=\ket{{}^+X,{}^+X}$, where $\ket{{}^{\pm}X}_i\equiv \frac{1}{\sqrt{2}}(\ket{\mathrm{g}}_i \pm e^{i\Delta t} \ket{\mathrm{e}}_i)$ are the $\pm$ eigenstates of $\sigma_x^{(i)}$.  
After the pulse, the molecules again evolve freely in time under the influence of Eq.~\ref{XXHamiltonian}.
%Since the dipole moments of the molecules are oscillating in time during this interaction, they average to zero in the lab frame and this naturally protects the phase accumulation from the influence of low-frequency electric fields from, for instance, the ion trap.
After a time $t_\mathrm{G} = \frac{\pi}{2J}$, the system has evolved to the entangled state $\ket{\psi^\prime} = \frac{1}{\sqrt{2}} e^{-\imath \pi/4} \left(\ket{{}^+X,{}^+X} + \imath \ket{{}^-X,{}^-X}\right)$.  Another fast $\frac{\pi}{2}$ pulse yields the singlet Bell state $\ket{\psi_\mathrm{f}} = \frac{1}{\sqrt{2}}e^{-\imath \pi/4}\left( \ket{\mathrm{e},\mathrm{e}}-\ket{\mathrm{g},\mathrm{g}}\right)$.
Choosing $t_G$ such that $\Delta t_G/\pi$ is an integer removes any detrimental effects from the counter rotating terms neglected in the rotating wave approximation that led to Eq.~\ref{XXHamiltonian}. 
This scheme is essentially a Ramsey sequence, and can be intuitively understood by considering the polarization induced by the first $\frac{\pi}{2}$-pulse.  
For instance, if microwaves that are linearly polarized along the molecular quantization axis are used to create the initial superposition, the resulting (time-dependent) dipole moments of the molecules are $\bra{{}^+X}\vec{\mathbf{d}}\ket{{}^+X} = d \cos(\Delta t)\hat{\mathbf{z}}$. 
Since these induced dipoles of the two molecules oscillate at the same frequency, there is a net dipole-dipole interaction. 
In the architecture of Fig.~\ref{fig:schematic}, this procedure could be accomplished by, for example, shuttling two ground state ions into the interaction region and subjecting them to fast, global $\pi/2$-pulses. 

Because the dipole moments average to zero in the lab frame for both of these implementations, these sequences naturally protect the phase accumulation from the influence of low-frequency electric fields arising from, for instance, the ion trap.
Likewise, as no external fields need to be applied during the gates, the phase accumulation is less sensitive to amplitude noise on the radiation than a continuously-driven gate. 
Using these sequences as basic building blocks, quantum gates can be engineered on the system.

%\section{Implementations}
For qubits defined on molecular ion rotational states, spontaneous emission does not significantly limit the gate fidelity.
However, there are other sources of decoherence to consider. 
Many of these effects have been considered extensively in Ref.~\cite{Schuster2011}. 
Therefore, we only briefly review the main features of environmental decoherence and show that many of the effects considered in Ref.~\cite{Schuster2011} can be suppressed if the the polarization axis is taken to be along the trap axial direction. 

If $\sigma^\pm$ radiation is used to create the superposition, the state is potentially magnetically sensitive and then typical decoherence sources for magnetically sensitive qubits apply, which can be presumably addressed with the normal techniques~\cite{Harty2014}. 
Further, the superposition states, e.g. $\left|^\pm X\right\rangle$, are not energy eigenstates in an electric field, therefore any electric field experienced by the ion can lead to decoherence. 

This electric field leads to decoherence in two main ways. 
First, electric fields along the direction of polarization ($\vec{\mathbf{E}}_\parallel$) cause a differential Stark shift between the energy eigenstates, leading to an extra superposition phase accumulation of $\phi = \int \mathrm{d}t\frac{2 d^2 E_{\parallel}^2}{\Delta}$.
For a trapped ion chain, ion collisions can be ignored and $\langle\vec{\mathbf{E}}_\parallel^2\rangle < \langle\vec{\mathbf{E}}_{\mathrm{trap}}^2\rangle$. 
Thus, for dipoles along a radial direction of the trap, in the limit of low Mathieu $q$ parameter~\cite{Chen2014}, the  accumulated phase is $\phi \lesssim \frac{d^2 V_{\mathrm{rf}}^2 k_\mathrm{B} T}{m \omega_r^2 r_o^4 \Delta}t$, where $\omega_r$ is the secular frequency and $r_o$ is the trap field radius. 
For dipoles along the axial direction of the trap $\phi \approx \frac{4d^2m\omega_a^2}{e^2}\left(\frac{k_B T}{\Delta}\right)$, where $\omega_a$ is the axial secular frequency.
As long as the ion secular motion is not significantly perturbed, traditional spin echo techniques should be able to mitigate this effect~\cite{Yan2013}.

Second, any electric field that is not along the original polarization direction ($\vec{\mathbf{E}}_\perp$) will induce population transfer from $\ket{\mathrm{g}}$ into the other Zeeman components of the $\left|\mathrm{e}\right\rangle$ manifold. 
The population transferred into these other states goes as $\mathscr{L} \approx d^2 E_\perp^2 \left|\int_0^t\mathrm{d}t^\prime e^{\imath\Delta t'}\right|^2 \leq \frac{4 d^2 \langle\vec{E}^2\rangle}{\Delta^2}.$
This result contrasts with what would be expected for a magnetic dipole. 
Because eigenstates of parity can possess a magnetic dipole, the magnetic dipole can fully reorient 
%from $\ket{\mathrm{e}}$ into the degenerate Zeeman substates of that level
at a rate of $\mu B$, as expected from degenerate perturbation theory.
For the electric dipole, reorientation only proceeds via population transfer from $\ket{\mathrm{g}}$ into the undesired Zeeman sublevels of the manifold that contains $\ket{\mathrm{e}}$ and is therefore suppressed by $\Delta$. 

In addition to the dephasing due to the trap fields sampled by the thermal motion of the ion, any stray electric field, $\vec{\mathbf{E}}_{\rm{s}}$, can shift the ion from the center of the trap resulting in $\langle\vec{\mathbf{E}}^2\rangle\lesssim 100\vec{\mathbf{E}}^2_{\rm{s}}$~\cite{Schuster2011}. Micromotion compensation techniques can control the residual field to the level of $|\vec{\mathbf{E}_{\rm{s}}}| \approx 0.1$~V/m~\cite{Gloger2015}, leading to another source of decoherence that is equivalent to the thermal decoherence rate at $\approx$1~mK. 

Similarly, the thermal motion of the ion also leads to gate errors as the value of $J$ changes with ion separation $r$. 
The dipole-dipole interaction strength is maximized and the thermal motion decoherence rate is minimized for dipoles oriented along the axial direction of a linear chain. 
In this case, taking $t_G = \frac{\pi}{4\left<J\right>}% = \frac{e^2}{4d^2m\omega_a^2}
$, the fidelity of the entangling gates is $\mathscr{F}_G = \cos^2\left(\frac{3\pi k_B T}{4}\left(\frac{4\pi^2\epsilon_o^2}{m e^4\omega_a^2}\right)^{1/3}\right)$. For $\omega_a = 2\pi\times10$~MHz and $T = 0.1$~K, $\mathscr{F}_G \geq 0.9999$ for the CO$^+$ molecule described later. 
During the gate time, the electric field noise described above leads to a secular frequency independent limit on the fidelity, $\mathscr{F}_E = \cos^2\left(\frac{k_B T}{2\Delta}\right)$, which for CO$^+$ at $T=0.1$~K is $\mathscr{F}_E \geq 0.9999$.

Last, the interaction between the molecular dipole moment and the trapping fields can off-resonantly couple the oscillation of the dipole to the ion motion.  This gives rise to a motion-mediated effective dipole-dipole interaction.  In the secular approximation for the trap, a perturbative estimate of this effect indicates that it can be approximated by Eq.~\ref{XXHamiltonian} times $\kappa_j\omega_j^2/(\Delta^2 - \omega_j^2)$ where $\kappa_j$ is a signed, dimensionless quantity of order unity that depends upon the details of the normal mode $j$.  This interaction is therefore suppressed unless one of the secular frequencies $\omega_j$ becomes comparable to the dipole oscillation frequency $\Delta$.  For a two-ion crystal, $\kappa_1$ and $\kappa_2$ have opposite signs, and the motion-mediated dipole-dipole interaction is suppressed even further.

\begin{figure}
\includegraphics[width=\textwidth]{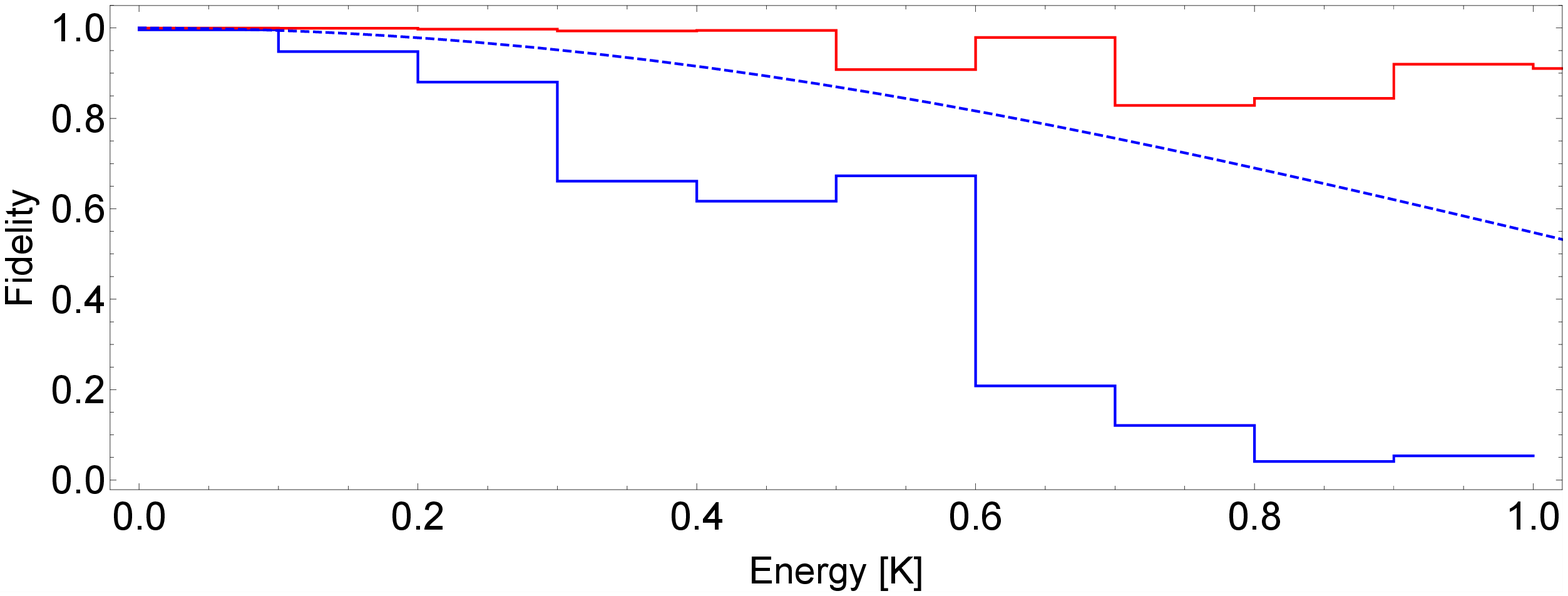}
\includegraphics[width=\textwidth]{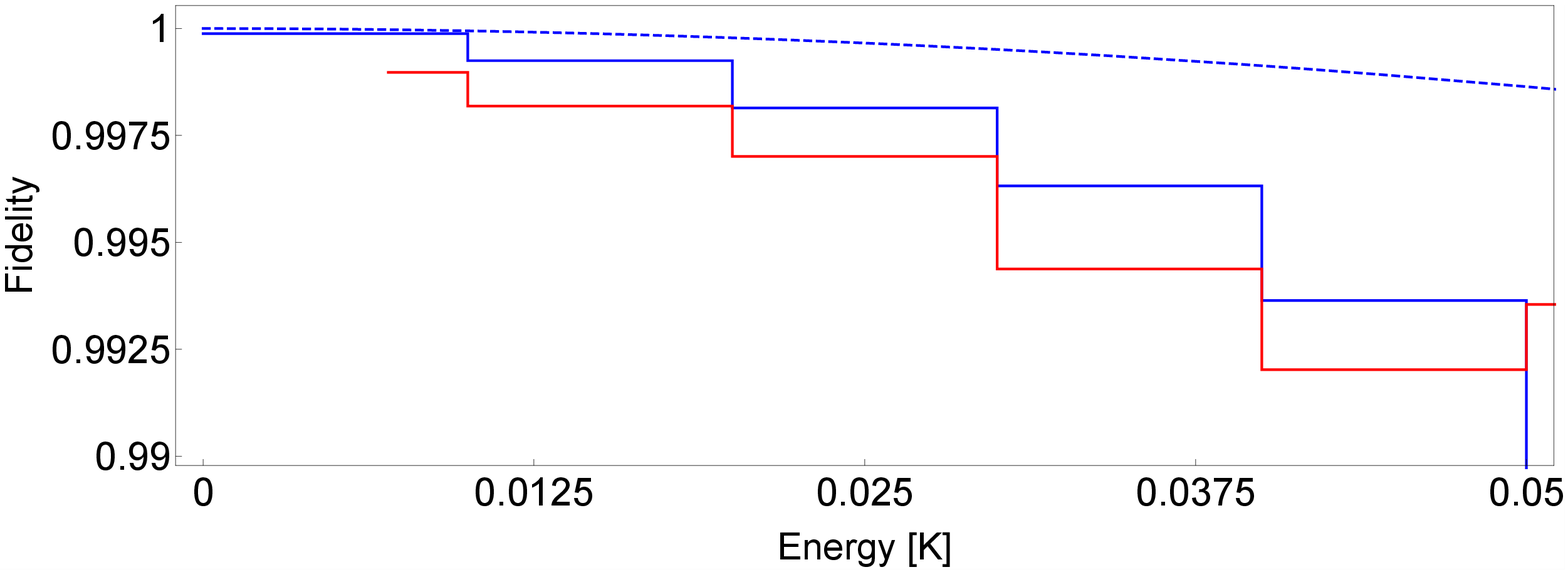}
\includegraphics[width=\textwidth]{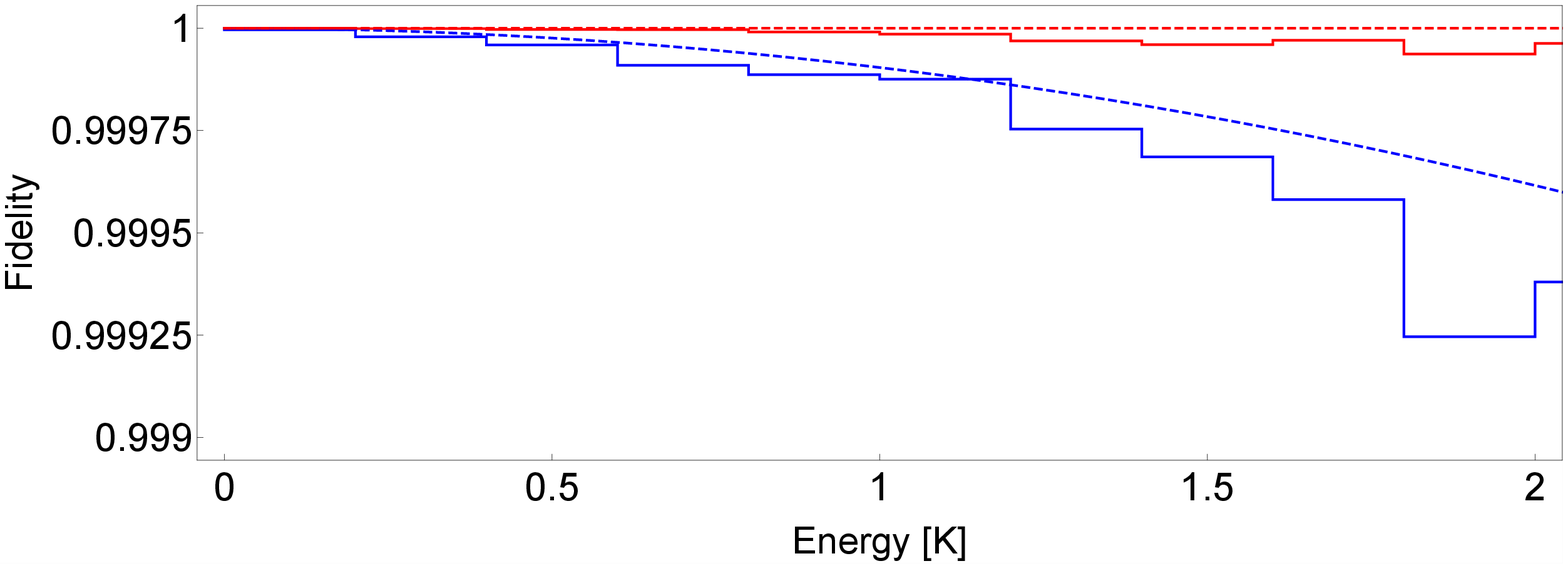}
\caption{The expected  fidelity for the procedure outlined here as a function of temperature for $100~\mu$s of evolution. Unless noted the molecular dipole moment is parallel to the trap radius. (Top) Results for BaCl$^+$ with (red bins) and without (blue bins) the use a spin echo pulse. The dashed blue line is the result estimated in the text. (Middle) The BaCl$^+$ result without echo is compared to the result when a stray electric field of 1~V/m is applied. (Bottom) The result for CO$^+$ without echo is shown (blue bins) next to the estimate from the text (dashed blue line). The red bins and line are for the case when the dipole is oriented along the trap axis.}
\label{fig:simresults}
\end{figure}

A numerical simulation was performed to verify these estimates for decoherence. 
For this simulation, the trajectory of two ions in a linear quadrupole trap 
%with $q = 0.2$ and secular frequency of 80~kHz (for BaCl+)
%with $q = 0.15$ and secular frequency of X~kHz (for CO+)
is solved classically for 300~$\mu$s. 
The concurrent evolution of the internal quantum states is found by integrating Schr\"odinger's equation using the classically determined environmental electric fields. 
The dipole-dipole interaction between the ions is ignored in order to focus on evolution due to decohering effects. 
In this simulation, a fictitious laser cooling force is first applied for a random amount of time, always less than 50~$\mu$s, to prepare the ions at different energy.
At $t=150$~$\mu$s, an approximate $\frac{\pi}{2}$ pulse is applied to drive population from $\ket{\mathrm{g}}$ into a superposition state. 
This state then evolves for 100~$\mu$s in the electric field of the ion trap and the other ion before the opposite, approximate $\frac{\pi}{2}$ pulse is applied. 
If no decoherence occurs during the evolution, the second pulse returns all the population to $\ket{\mathrm{g}}$. 
In these simulations, the hyperfine structure of the molecule is ignored. 
This constraint is relaxed later, when a practical implementation is considered.

The results are binned and plotted in Fig.~\ref{fig:simresults}, where the projection $\left|\left\langle\mathrm{g}|\psi\right\rangle\right|^2$ after the sequence is shown versus initial energy of the ions. 
In panel Fig.~\ref{fig:simresults}(a) the blue bins are the results for BaCl$^+$ in a trap similar to that employed in Ref.~\cite{Puri2017} with the dipoles oriented along the trap radius, the blue line is the result predicted by the estimates above, $\mathscr{F} = \cos^2{\left(\frac{\phi}{2}\right)} $, and the red bins show the result when a spin echo $\pi$-pulse is applied at the middle of the free evolution time (150~$\mu$s). 
Fig.~\ref{fig:simresults}(b) compares the result for BaCl$^+$ with perfect compensation of stray electric fields (blue bins) to that with an uncompensated 1~V/m stray electric field along a radial direction. 
Fig.~\ref{fig:simresults}(c) shows the results when BaCl$^+$ is replaced by CO$^+$. The blue bin and lines are the simulation results and prediction, respectively, for the case where the dipoles are oriented along the trap radius; the red bins and line are the same, but for axial dipole alignment. 

One of the most striking results of this simulation is that the fidelity is, when compared to atomic ion quantum information processing, relatively insensitive to the ion temperature. This coupled with the fact that the entanglement is not generated through motional trap modes, means that a molecular ion quantum computer does not require ground state motional cooling. This implies the technique is relatively insensitive to anomalous heating, which is one of the main roadblocks to ion trap quantum computing scalability~\cite{Sedlacek2018}. 

%\subsection{A practical molecular ion implementation} 
While these decoherence times are sufficient for high-fidelity entangling operations, they are significantly shorter than what is achievable for a qubit defined on a magnetically insensitive hyperfine transitions. 
Therefore, it is advantageous to use a polar molecular ion with hyperfine structure and employ the dipole-dipole interaction to achieve entanglement. 
In this manner, long qubit storage times are possible. 

\begin{figure}
\includegraphics[width=\textwidth]{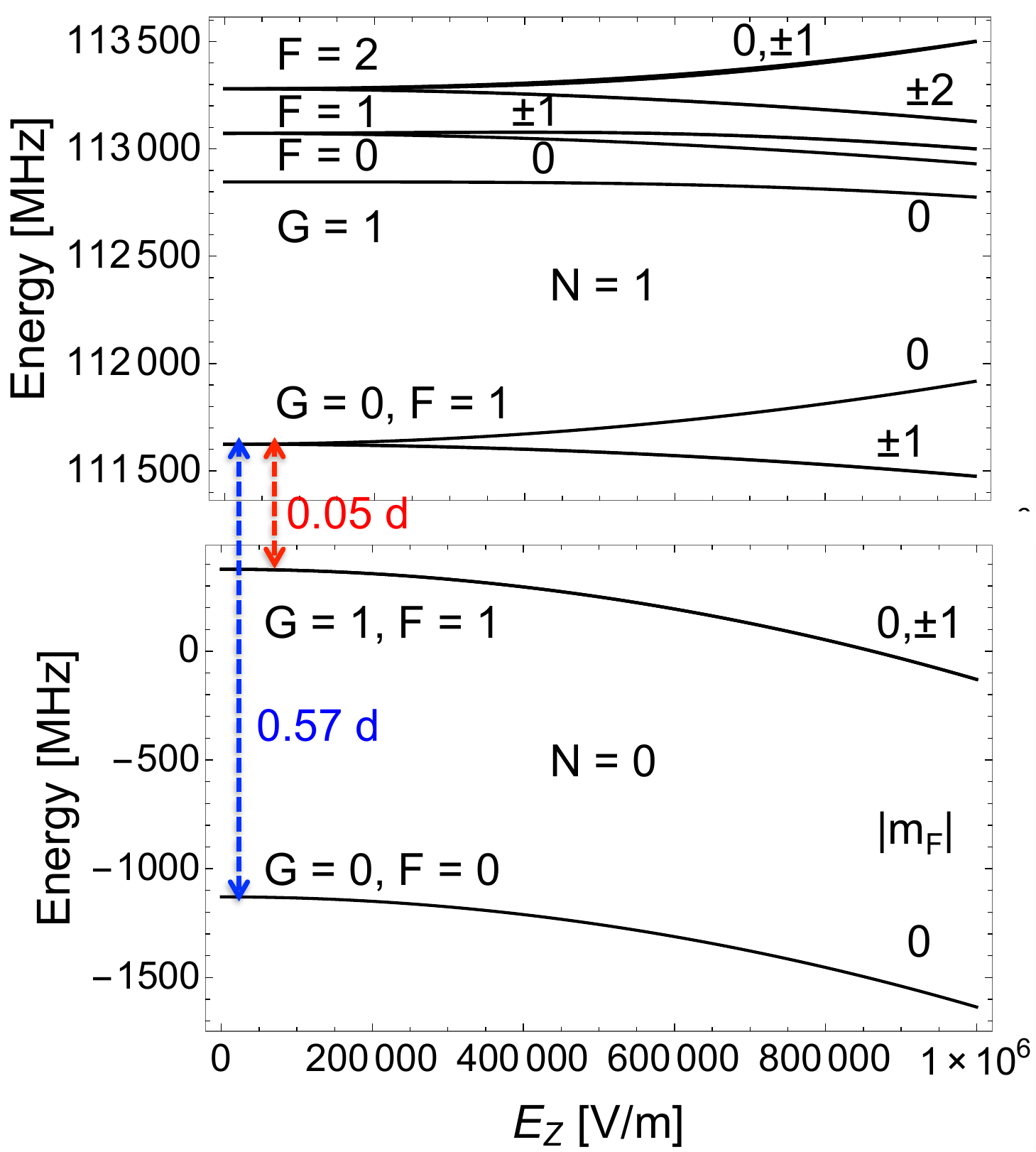}
\caption{The Stark shift of the lowest two rotational levels of $^{13}$CO$^+$. Relevant transitions dipole moments are shown ($d = 1.025ea_o$~\cite{Thompson2004}). The $|2\rangle \leftrightarrow |1\rangle$ transition (red arrow, see text) is subject to the $m_F = 0 \not\leftrightarrow m_F = 0$ selection rule.}
\label{fig:co}
\end{figure}

As a concrete, and particularly promising, example we consider the $^{13}$CO$^+$ molecule, whose energy structure is shown in Fig.~\ref{fig:co}. 
This molecule possesses a $^2\Sigma$ electronic ground state, whose rotational structure is split by both hyperfine and spin-rotation interactions. 
The hyperfine interaction is larger than the spin-rotation, leading to Hund's case b$_{\beta S}$ coupling, where the nuclear spin $\vec{I}$ and electronic spin $\vec{S}$ are coupled to form $\vec{G} = \vec{I} + \vec{S}$, which is coupled with the rotational angular momentum $\vec{N}$ to form the total angular momentum $\vec{F} = \vec{N} + \vec{G}$. 
In the absence of the spin-rotation interaction, $\Delta G \neq 0$ transitions are forbidden, however, spin-rotation mixes states with different $G$ but the same $F$, relaxing this selection rule as denoted by the red arrow in Fig~\ref{fig:co}.

Defining a hyperfine (storage) qubit on $\ket{\mathrm{g}} \equiv \ket{N = 0, F = 0}$ and $\ket{\mathrm{g}^\prime} \equiv \ket{N = 0, F = 1, m_F = 0}$, split by $\delta = 2\pi\times1511.5$~MHz states has several advantages. 
First, as demonstrated in $^{171}$Yb$^+$, these so-called clock-state qubits can have magnetically-limited coherence times exceeding 10 minutes~\cite{Wang2017}. 
Second, the differential Stark shift between these two states is $-1.4\times10^{-7} E^2$~[Hz/(V/m)$^2$], meaning coherence times in excess of 10$^4$~s are achievable at $\sim$1~mK. 
Third, site-specific single qubit rotations can be achieved using Raman lasers or near-field microwave gates. 
Fourth, using the Purcell effect from a tunable, coplanar waveguide, as described in Ref.~\cite{Schuster2011} and sketched in Fig.~\ref{fig:schematic}, at cryogenic temperatures optical pumping on the transition denoted by the red arrow in Fig~\ref{fig:co} provides a means for high-fidelity preparation of $\ket{\mathrm{g}}$.

Entanglement of two $^{13}$CO$^+$ molecules via the dipole-dipole interaction is straightforward. 
For two molecules initially in $\ket{\psi_o} = \frac{1}{\sqrt{2}}\left(\ket{\mathrm{g}}+\ket{\mathrm{g}^\prime}\right)_{(1)} \ket{\mathrm{g}}_{(2)}$ a global $\frac{\pi}{2}$-pulse on the $\ket{\mathrm{g}}\leftrightarrow \ket{\mathrm{e}} \equiv \ket{N = 1, G = 0, F = 1} \leftrightarrow |0\rangle $ transition, prepares $\ket{\psi} = \frac{1}{\sqrt{2}}\left(\ket{^+X,^+X}+\ket{\mathrm{g}^\prime,^+X}\right)$. After a time $t_\mathrm{G} = \pi/J$, this state evolves into $\ket{\psi} = \frac{1}{\sqrt{2}}\left(\ket{^-X,^-X}+e^{-\imath\delta t_\mathrm{G}}\ket{\mathrm{g}^\prime,^+X}\right)$. A second $\frac{\pi}{2}$-pulse, opposite to the first, produces $\ket{\psi} = \frac{1}{\sqrt{2}}\left(\ket{\mathrm{e},\mathrm{e}}+e^{-\imath\delta t_\mathrm{G}}\ket{\mathrm{g}^\prime,\mathrm{g}}\right)$, which can, for example, be turned into $\ket{\psi} = \frac{1}{\sqrt{2}}\left(\ket{\mathrm{g},\mathrm{g}}+e^{-\imath\delta t_\mathrm{G}}\ket{\mathrm{g}^\prime,\mathrm{g}^\prime}\right)$ with $\pi$-pulses on the relevant transitions.

In summary, we have described a procedure for effecting a net dipole-dipole interaction in the absence of a laboratory frame dipole moment to entangle two trapped molecular ions with high fidelity ($\geq$~0.9999) without the need for cooling to the motional ground state. 
We have also described the use of advantageous hyperfine structure to achieve electric-field-noise-limited coherence times $>10^4$~s in a standard ion trap environment. 
This technique may be useful in the growing field of ultracold molecular ion trapping~\cite{Tomza2017,Hudson2016} and could aid efforts to scale trapped ion quantum computing to large numbers of qubits. 
It should also be extendable to many qubit systems, where it could allow, for example, quantum simulation of spin models with novel geometries~\cite{Yoshimura2016}.

We thank David Patterson, Alexey Gorshkov, and Paul Hamilton for useful discussions.

\bibliography{no_DipoleQC}

\end{document}